\documentclass[doublecol]{epl2}
\usepackage{graphicx}

\title{Nonergodicity and Central Limit Behavior for Long-range  Hamiltonians}
\shorttitle{Title} 

\author{A. Pluchino\inst{1} \and A. Rapisarda\inst{1} \and C. Tsallis\inst{2,3}}

\institute{
  \inst{1} Dipartimento di Fisica e Astronomia, Universit\`a di
Catania, and INFN sezione di Catania, - Via S. Sofia 64, I-95123
Catania, Italy\\
  \inst{2} Centro Brasileiro de Pesquisas Fisicas, - Rua Xavier Sigaud 150, 22290-180 Rio de Janeiro-RJ, Brazil \\
  \inst{3} Santa Fe Institute,  1399 Hyde Park Road, Santa Fe, NM 87501, USA
}

 { Complex systems}

\abstract{
We present  a molecular dynamics test of the  Central Limit Theorem (CLT)
in a paradigmatic long-range-interacting  many-body classical Hamiltonian system, the
HMF model. We  calculate sums of velocities at equidistant times  along  deterministic  trajectories for
different sizes and energy densities. We  show  that,
when the system is in a chaotic regime (specifically, at thermal equilibrium), ergodicity is essentially
verified, and the Pdfs  of the sums  appear to be Gaussians, consistently with the standard CLT.
When the system  is, instead,  only weakly chaotic (specifically, along longstanding metastable
Quasi-Stationary States),  nonergodicity (i.e., discrepant ensemble and time averages) is observed,
and robust $q$-Gaussian attractors emerge, consistently with recently proved generalizations of the CLT.
}

\begin{document}

\maketitle

\section{Introduction}

During recent years  there has been an increasing  interest  in generalizations of the
Central Limit Theorem  (CLT). This theorem -- so called because of its central position in theory of probabilities --
has ubiquitous and important applications in several fields.
It essentially states  that a  (conveniently scaled) sum  of $n\rightarrow \infty$ independent (or nearly independent) 
random variables with finite variance has a  Gaussian distribution.
Understandingly, this theorem is not
applicable to those complex systems where long-range correlations are the rule, such as those
addressed by nonextensive statistical mechanics \cite{tsallis0,tsallis}. Therefore, several papers  \cite{tsallisMilan,umarov,tirnakli,baldovin,comment,hilhorst,plastino,thistleton} have
recently discussed  extensions of the CLT and their corresponding attractors.
In this  paper, following   \cite{tirnakli,comment}, we  present  several   numerical
simulations  for  a  long-range Hamiltonian system, namely the Hamiltonian Mean Field (HMF) model.  This  model  is  a
paradigmatic one for classical Hamiltonian systems with long-range interactions which has been
intensively studied  in the last decade (see, for example,   \cite{hmf0,hmf1,hmf2,pluchino,ep1,glass2,giansanti,tamarit,chavanis,antoniazzi,morita,comment}, and references therein).
In \cite{tirnakli} it was   shown  that
the probability density of rescaled sums of iterates of deterministic dynamical systems (e.g., the logistic map) 
at the edge of chaos (where the  Lyapunov exponent vanishes) violates  the CLT.
Here we  study rescaled sums of velocities considered  along deterministic trajectories in the HMF model.
It is well known that, in this model, a wide class of out-of-equilibrium initial conditions induce a violent relaxation
followed by a metastable regime characterized by nearly vanishing (strictly vanishing in the thermodynamic limit) 
Lyapunov exponents,  and glassy dynamics \cite{pluchino,ep1,glass2}.
We exhibit that correlations and nonergodicity created along these Quasi-Stationary States (QSS)
can be so strong that, when summing the velocities calculated during the deterministic trajectories of single rotors
at fixed  intervals of time, the standard CLT is no longer applicable.
In fact, along the QSS, $q$-Gaussian  Pdfs emerge as attractors instead  of simple
Gaussian Pdfs, consistently with the recently  advanced
$q$-generalized  CLT  \cite{umarov,tirnakli,plastino}, and ensemble averages  are different from time averages.

\section{Numerical  simulations}

The HMF model describes a system of $N$ fully-coupled classical inertial
XY spins (rotors)
$\stackrel{\vector(1,0){8}}{s_i} = (cos~\theta_i,sin~\theta_i)~,~i=1,...,N,
$with unitary module and mass \cite{hmf0,hmf1}.
These spins can also be thought as particles rotating
on the  unit circle.
The  Hamiltonian is given by
\begin{equation}
\label{hamiltonian}
        H
= \sum_{i=1}^N  {{p_i}^2 \over 2} +
  { 1\over{2N}} \sum_{i,j=1}^N  [1-cos( \theta_i -\theta_j)]~~,
\label{eq.2}
\end{equation}
where ${\theta_i}$ ($0 < \theta_i \le 2 \pi$) is the angle
and $p_i$ the conjugate variable representing the rotational
velocity of spin $i$.
\\
The equilibrium solution of the model in the canonical ensemble
predicts a second order phase transition from a high
temperature paramagnetic  phase to a low temperature
ferromagnetic  one \cite{hmf0}.
The critical temperature is $T_c=0.5$ and corresponds to
a critical energy per particle $U_c = E_c /N =0.75$.
The order parameter of this phase transition is the modulus of
the {\it average magnetization} per spin defined as:
$M = (1 / N) | \sum_{i=1}^N
\stackrel{\vector(1,0){8}}{s_i} | ~~$.
Above $T_c$,  the spins point in different directions and $M \sim 0$.
Below $T_c$, most spins
are aligned (the rotators are trapped in a single cluster) and $M \neq0$.
The out-of equilibrium dynamics of the  model is also very interesting. In a range
of energy densities between $U\in[0.5,0.75]$, special  initial conditions called
\textit{water-bag}  (characterized by initial magnetization $M_0=1$ and uniform distribution of the momenta)
drive the system, after a violent relaxation,
towards  metastable QSS. The latter slowly decay towards equilibrium with a lifetime which diverges
like a power of the system size $N$ \cite{hmf2,pluchino,ep1}.

In this section we simulate the dynamical evolution of several HMF systems with different
sizes and at different energy densities,
in order to explore their behavior either inside or outside the QSS regime.
For each of them, following the prescription of the CLT, we construct probability density functions of quantities expressed as a finite sum of stochastic variables. But in this case, following the procedure adopted in ref.\cite{tirnakli} for  the  logistic  map, we will select these variables along the deterministics time evolutions of the N rotors.
More formally, we study the Pdf of the quantity $y$ defined as
 \begin{equation}
  y_j=\frac{1}{\sqrt{n}}\sum_{i=1}^n  (p_j(i) - <p_j>)  ~~~ for  ~~~ j=1,2,...,N ~~,
\label{eq.3}
\end{equation}
where  $p_j(i)$, with $i=1,2,...,n$,  are the  velocities of the $jth$-rotor taken at fixed intervals of time $\delta$ along the same trajectory. The latter are  obtained integrating the HMF equations of motions (see \cite{pluchino} for details  about these equations and the integration algorithm adopted
). The quantity  $<p_j>=(1 / n)\sum_{i=1}^n p_j(i)$ is the  average of the $p_j(i)$'s over the single  trajectory.
The product $\delta \times n$ gives the total simulation time. 
Note that the variables $y$'s are proportional to the \textit{time average} of the velocities
along the single  rotor  trajectories. In the following we will distinguish this kind of average, i.e. \textit{time average},  from the standard  \textit{ensemble average}, where the average  of the velocities  of  the $N$ rotators  is calculated a\textit{t a given fixed time } and over many different realizations of the dynamics.  The latter can also be obtained from eq.(\ref{eq.3}) considering the $y$'s variables with $n=1$ and $<p_j>=0$.
\begin{figure}
\begin{center}
\includegraphics [scale=0.23] 
{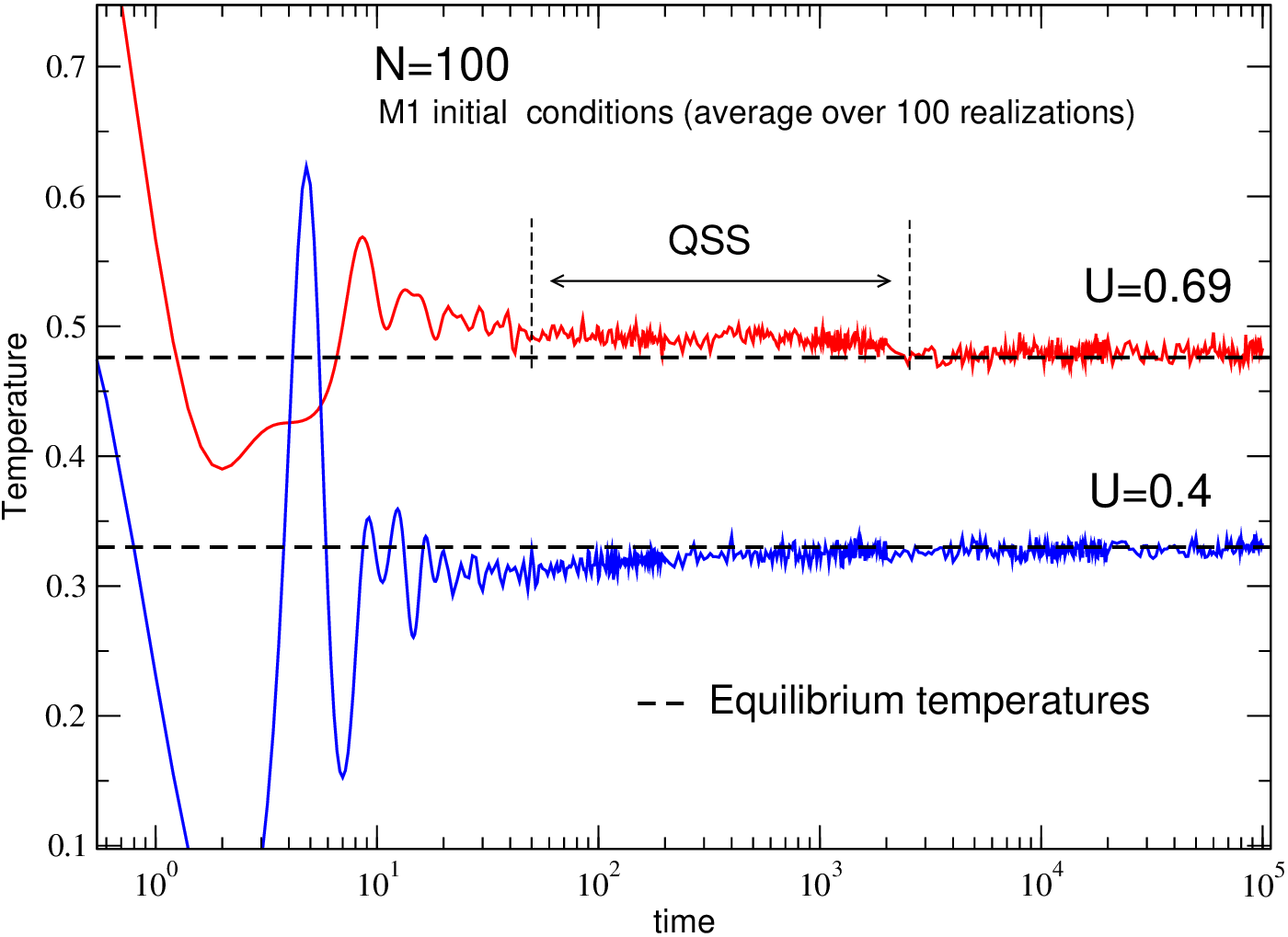}
\caption{Temperature time evolution for the HMF system,
with N=100 and M1 initial conditions, for $U=0.69$  and
for $U=0.4$. The presence of a QSS regime is visible
only in the $U=0.69$ case, although a transient regime exist also
for $U=0.4.$ 
} 
\label{fig1}
\end{center}
\end{figure}
%
In general, although the standard CLT predicts a Gaussian shape for sum of $n$ independent stochastic values strictly when $n\rightarrow \infty$, in practice a finite sum converges quite soon to the Gaussian shape and this, in absence of correlations, is certainly true at least for the central part of the distribution \cite{mantegna}.
Typically we will use in this section a sum of $n=50$ values of velocities along the deterministic trajectories for each of the $N$ rotors of the HMF system, though larger values of $n$ were also considered.

In the following we will show that, if correlations among velocities are strong enough and the system is weakly chaotic, CLT predictions are not verified and, consistently with  recent  generalizations of the  CLT, $q$-Gaussians appear \cite{tsallisMilan,umarov,tirnakli}.
The latter are  a  generalization of Gaussians which emerge in the context of nonextensive statistical mechanics \cite{tsallis0,tsallis} and are defined  as
\begin{equation}
 G_q(x)=A (1-(1-q) \beta x^2 )^{1/{1-q}}  ~~,
\end{equation}
being $q$ the so-called \textit{entropic index} (for $q=1$ one recovers the usual Gaussian) , $\beta$ another suitable parameter (characterizing the width of the distribution), and $A$ a normalization constant (see also ref. \cite{thistleton} for a  simple and general way to generate them).
In particular we will show in this section that:

(i) \textit{at equilibrium}, when correlations are weak and the system is strongly chaotic (hence ergodic) standard CLT is verified, and time average coincides with ensemble average (both corresponding Pdfs are Gaussians, either in the limit $n\rightarrow \infty$ or $\delta\rightarrow \infty$);

(ii) \textit{in the QSS regime}, where velocities are strongly correlated and the system is weakly chaotic and nonergodic, the standard CLT is no longer applicable, and  $q$-Gaussian attractors replace the Gaussian ones; in this regime ensemble averages do not agree with time averages.

For all the present simulations, water-bag initial conditions with initial magnetizazion $M_0=1$, usually referred as M1, will be used. In general, several different realizations of the initial conditions will be performed also for the time average Pdfs case, but only in order to have a good statistics for small values of $N$ (for N=50000, on the contrary, only one realization has been used: see fig.\ref{fig7}(b)). Finally, to allow a correct comparison with standard Gaussians (represented as dashed lines in all the figures) and $q$-Gaussians (represented as full lines), the Pdf curves were always normalized to unit area and unit variance, by subtracting from the $y$'s their average  $<y>$ and dividing by the correspondent standard deviation $\sigma$ (hence, the traditional $\sqrt{n}$ scaling adopted in Eq. (2) is in fact irrelevant).
%
\begin{figure}
\begin{center}
\includegraphics[scale=0.17]{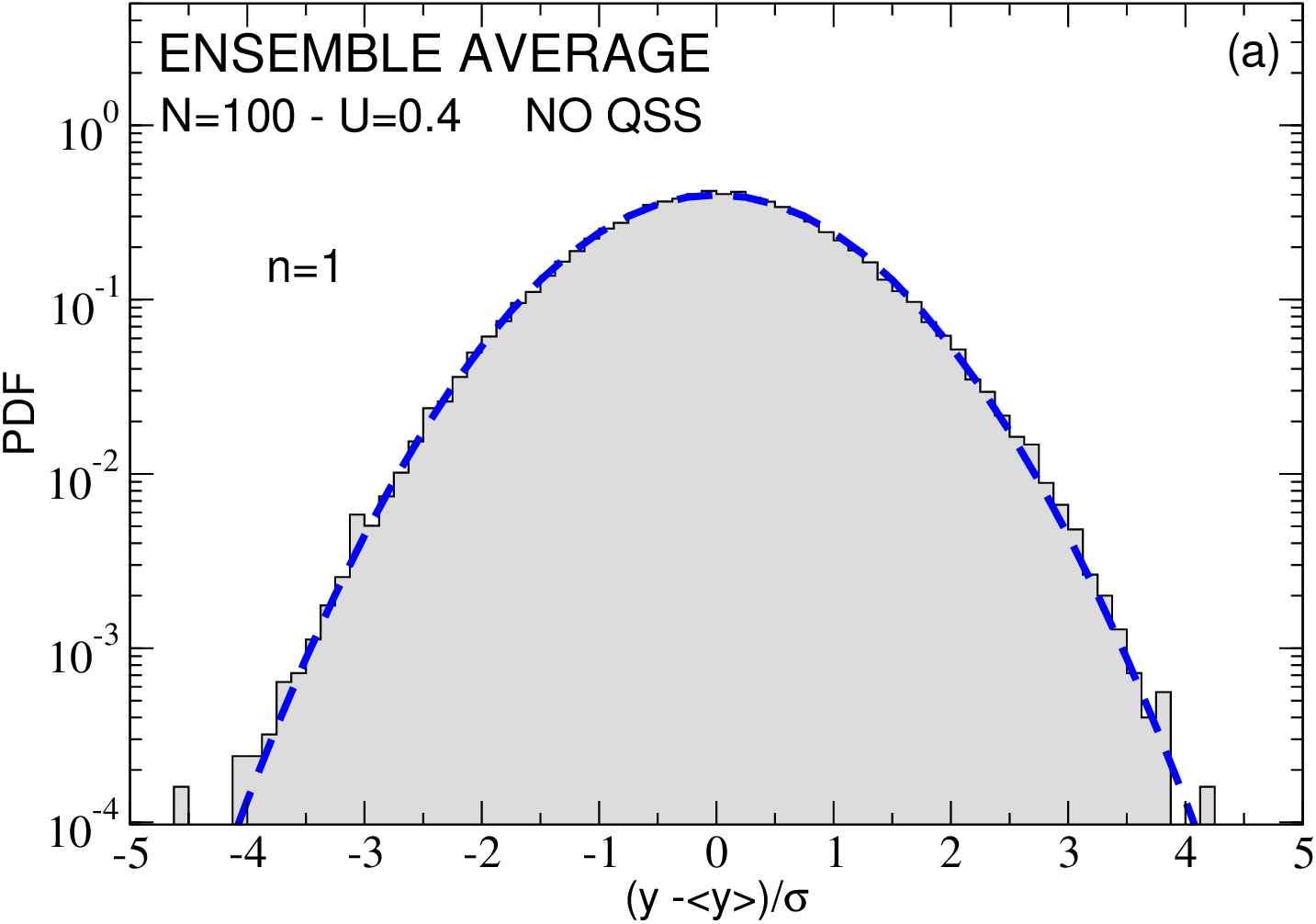}
\includegraphics[scale=0.17]{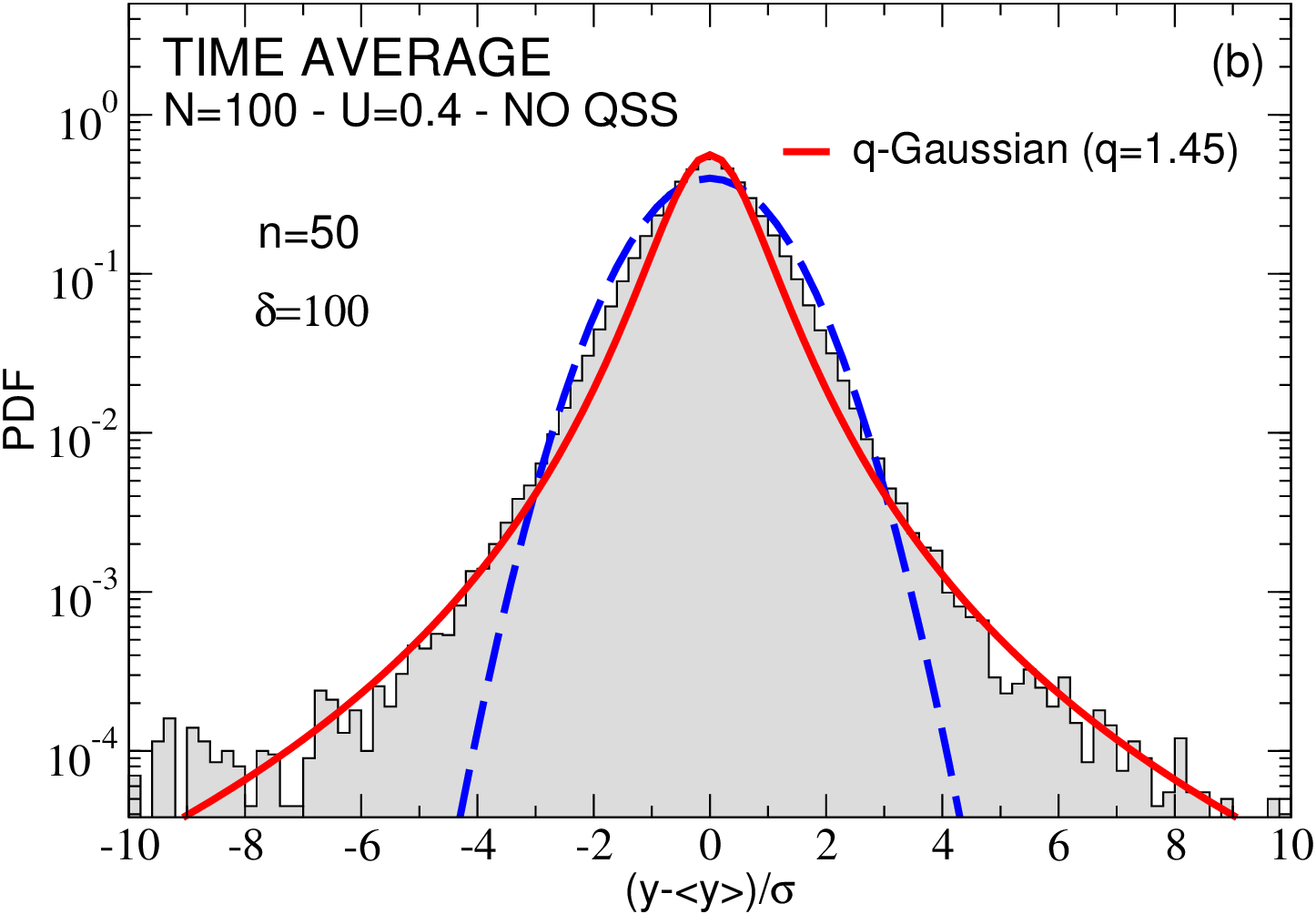}
\includegraphics[scale=0.17]{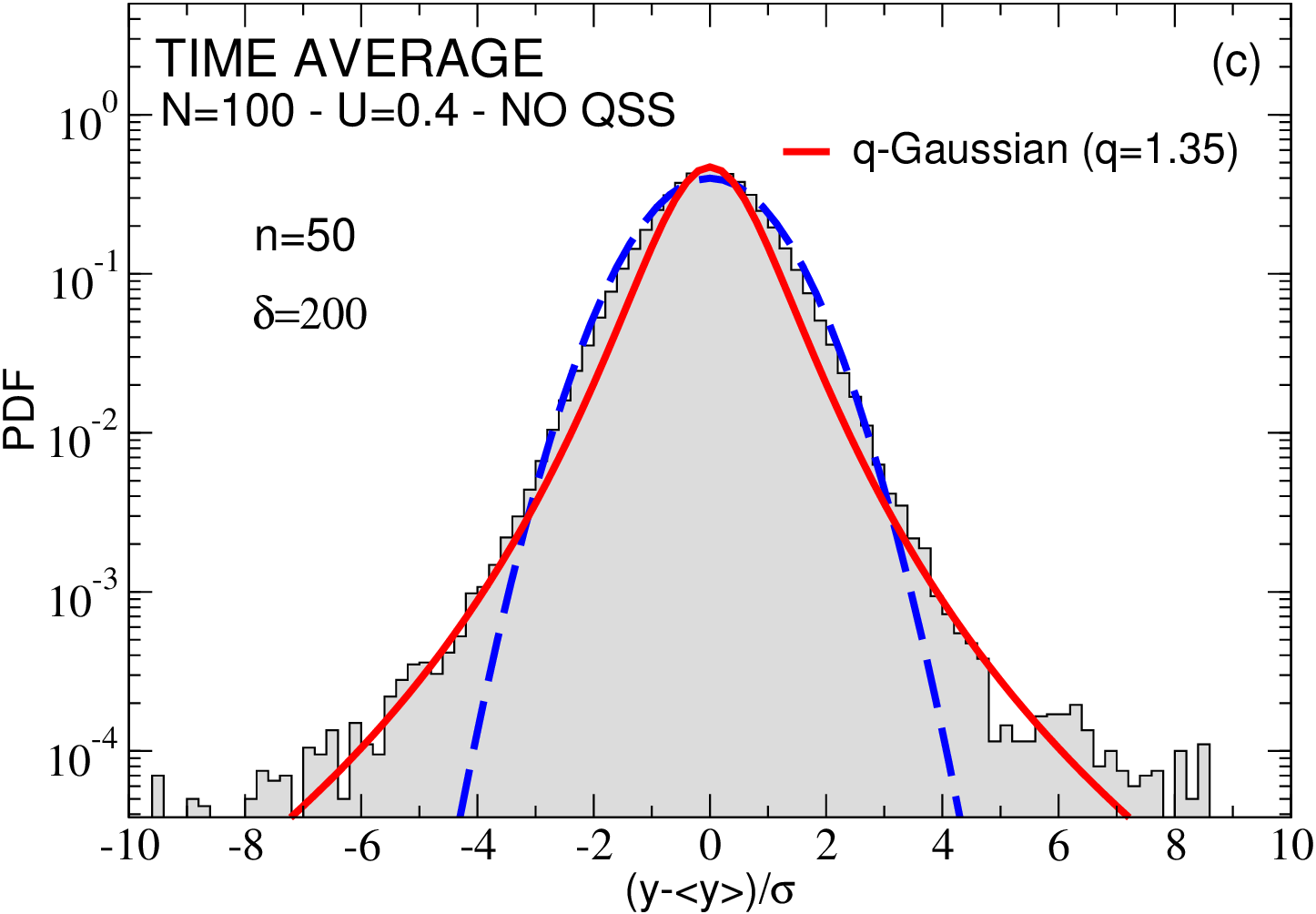}
\includegraphics[scale=0.17]{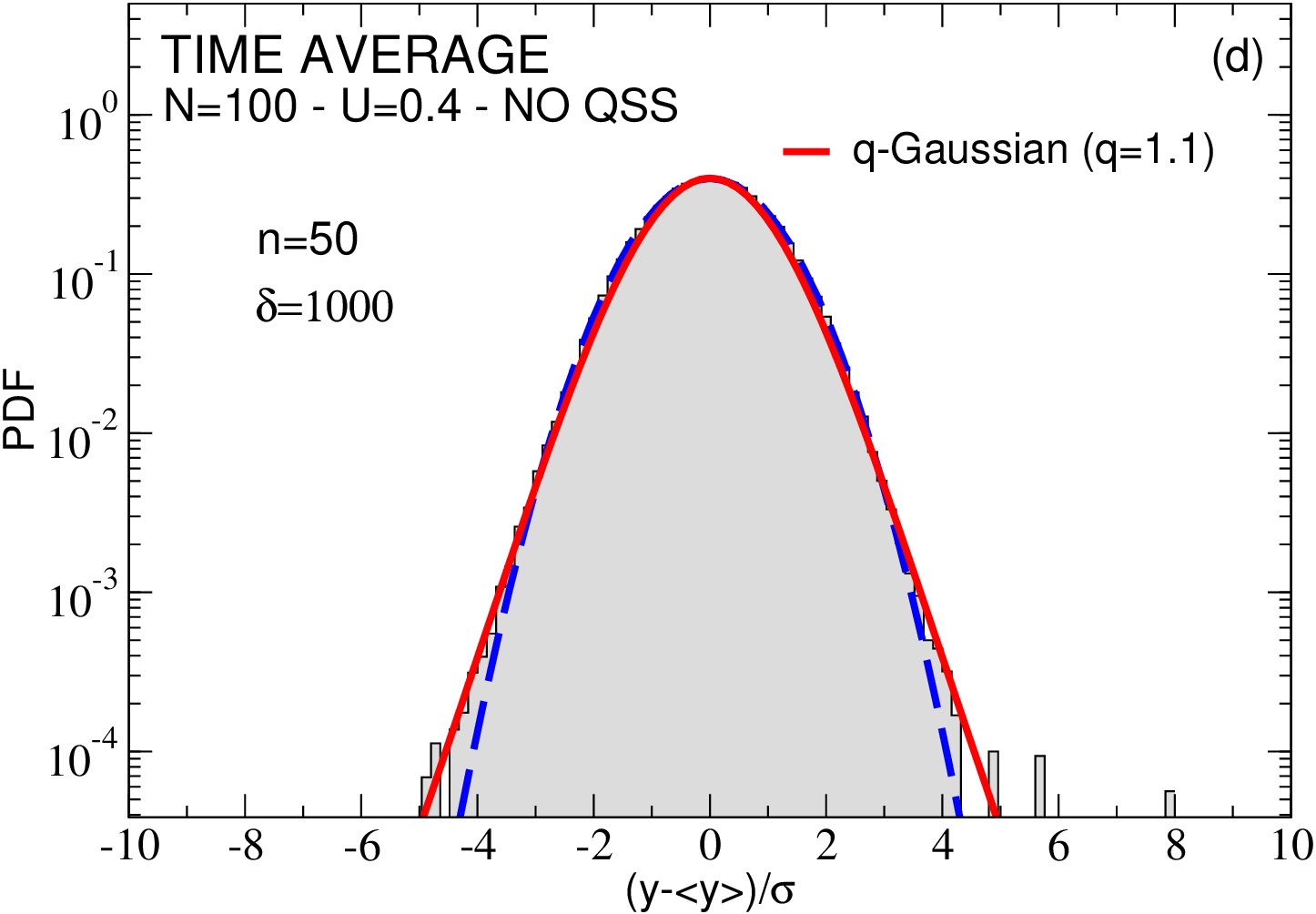}
\caption{Numerical simulations for the HMF model with N=100, U=0.4 and M1 initial conditions. No QSS are present for this energy value. (a) We plot here the  Pdf  of the single rotor velocities at the time t=40000 (ensemble average over 1000 realizations), i.e.  we plot the (normalized) variable  $y$  defined as in eq.(\ref{eq.3}) with $n=1$. The shape is Gaussian since the system is at equilibrium.
In the other figures (b), (c) and (d) we plot the time average Pdfs for the  normalized variable $y$, with
$n=50$ but with different  time intervals  ($\delta$=100,200 and 1000), calculated over an increasing simulation time after a transient of 40000 time  units. An average over 1000 different realizations of the initial conditions was also considered in order to have a good statistics. Even if we are at equilibrium, it is evident a strong dependence of the entropic index $q$ of the $q$-Gaussian fitting curve  on  the time interval $\delta$ adopted. Anyway, a time interval $\delta=1000$ is already sufficient to obtain a Gaussian-shaped Pdf. See text for further details.}
\label{fig2}
\end{center}
\end{figure}

\subsection{The case N=100}

We start the discussion of the numerical simulations for the  HMF  model considering a size  $N=100$  and two different  energy densities, $U=0.4$  and $U=0.69$.  In the first case  no QSS exist,  while  in the second  case  QSS  characterize the  out-of-equilibrium dynamics and correlations formed  during the first part of the dynamics decay slowly while the system relaxes towards  equilibrium \cite{pluchino,ep1}. With $N=100$ this relaxation takes however a reasonable amount of time steps, thus one can easily study also the equilibrium regime. The situation is illustrated in fig. \ref{fig1},
where we show  the  time evolution of the  temperature - calculated as twice the average kinetic energy per particle -    for the two energy densities considered, starting from $M_0=1$ initial conditions. As expected QSS are clearly visible only in the case $U=0.69$, although a small transient regime exists also for the case $U=0.4$ \cite{pluchino}. 

\subsection{N=100 and U=0.4}
Here  we discuss  numerical simulations for  the HMF model with size $N=100$  and $U=0.4$.
In this case it has been  shown in the past that the equilibrium regime is reached quite fast and is
characterized by a very chaotic dynamics \cite{hmf0,hmf1}.

In fig. \ref{fig2} a transient time of $40000$ units has been performed before the calculations, so that  the  equilibrium  is fully reached (see  fig.\ref{fig1}).  In (a) we consider the ensemble average of the velocities, i.e. the $y$ variables defined as in (\ref{eq.3}) with $n=1$, at $t=40000$ and taking $1000$ different realizations of the initial conditions (events). The Pdf compares very well with the Gaussian curve (dashed line), as expected at equilibrium.
On the other hand, we consider in (b), (c) and (d) the Pdfs for the  variable  $y$ with  $n=50$ and
with different time intervals $\delta$ over an increasing simulation time at equilibrium.
As previously explained, this procedure corresponds to performing a time average along the trajectory for all the rotors of the system. In this case  only the central part of the curve  exhibits  a  Gaussian  shape. On the other hand, Pdfs have long fat tails which can be very well reproduced   with $q$-Gaussians (full lines).  If one  increases   the  time interval $\delta$   going from $\delta=100$ (b), to $\delta=200$ (c) and finally   to $\delta=1000$ (d), the  tails tend to disappear, the entropic index $q$ of the $q$-Gaussians decreases from $q=1.45\pm0.05$ towards $q=1$ and the Pdf tends to the standard Gaussian. This means  that, as expected, summed velocities are less and less correlated as $\delta$ increases (see also ref.\cite{tirnakli}) and  therefore  the  assumptions of the  CLT are   satisfied as well as its prediction.
Notice that $n=50$ terms and a time interval  $\delta=1000$ are sufficiently large to reach a Gaussian-shaped Pdf.  This situation  reminds  similar observations  in the analysis  of returns in financial markets \cite{mantegna}, 
or in turbulence \cite{beck}.

\begin{figure}
\begin{center}
 \includegraphics[scale=0.178]{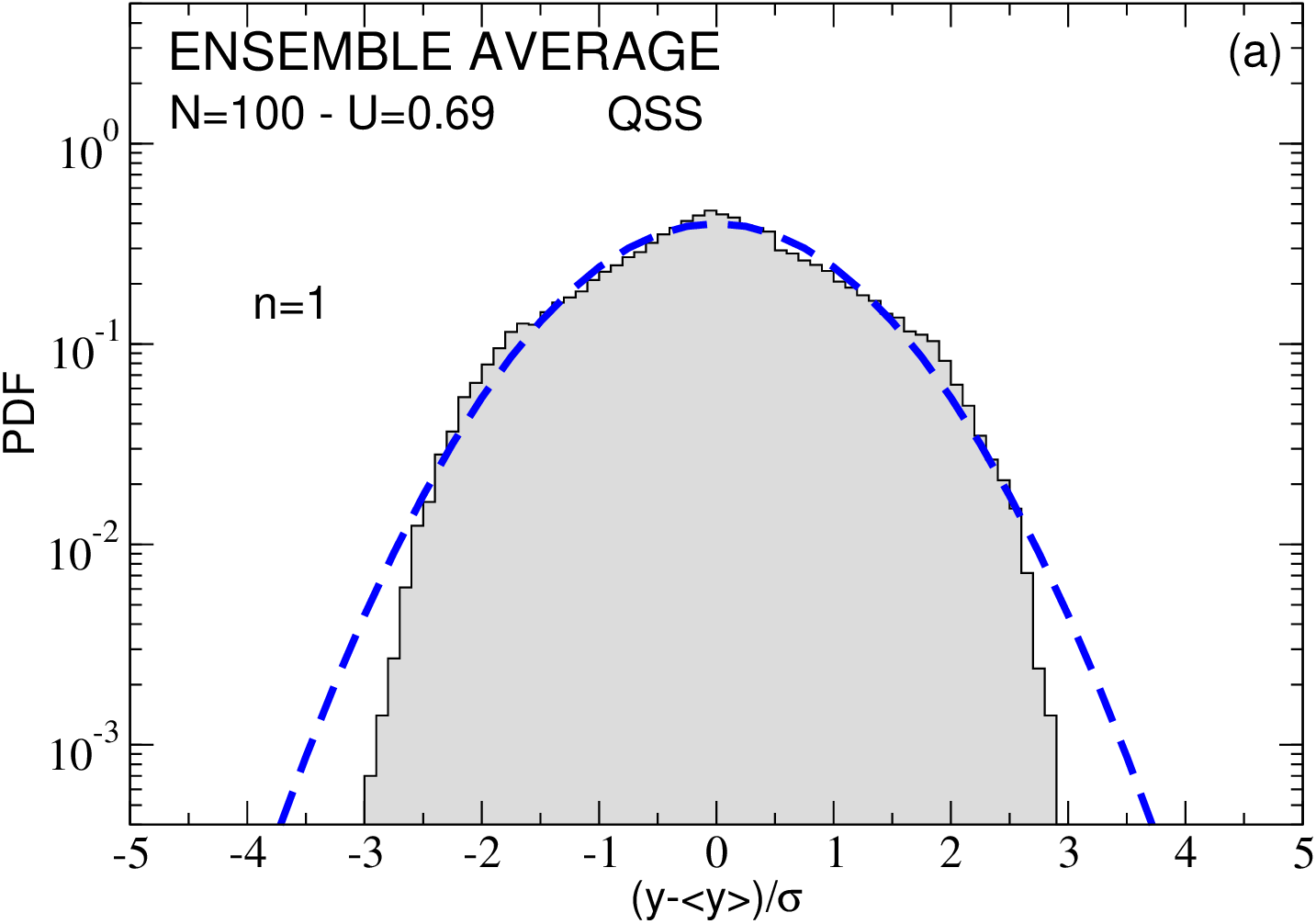}
\includegraphics[scale=0.178]{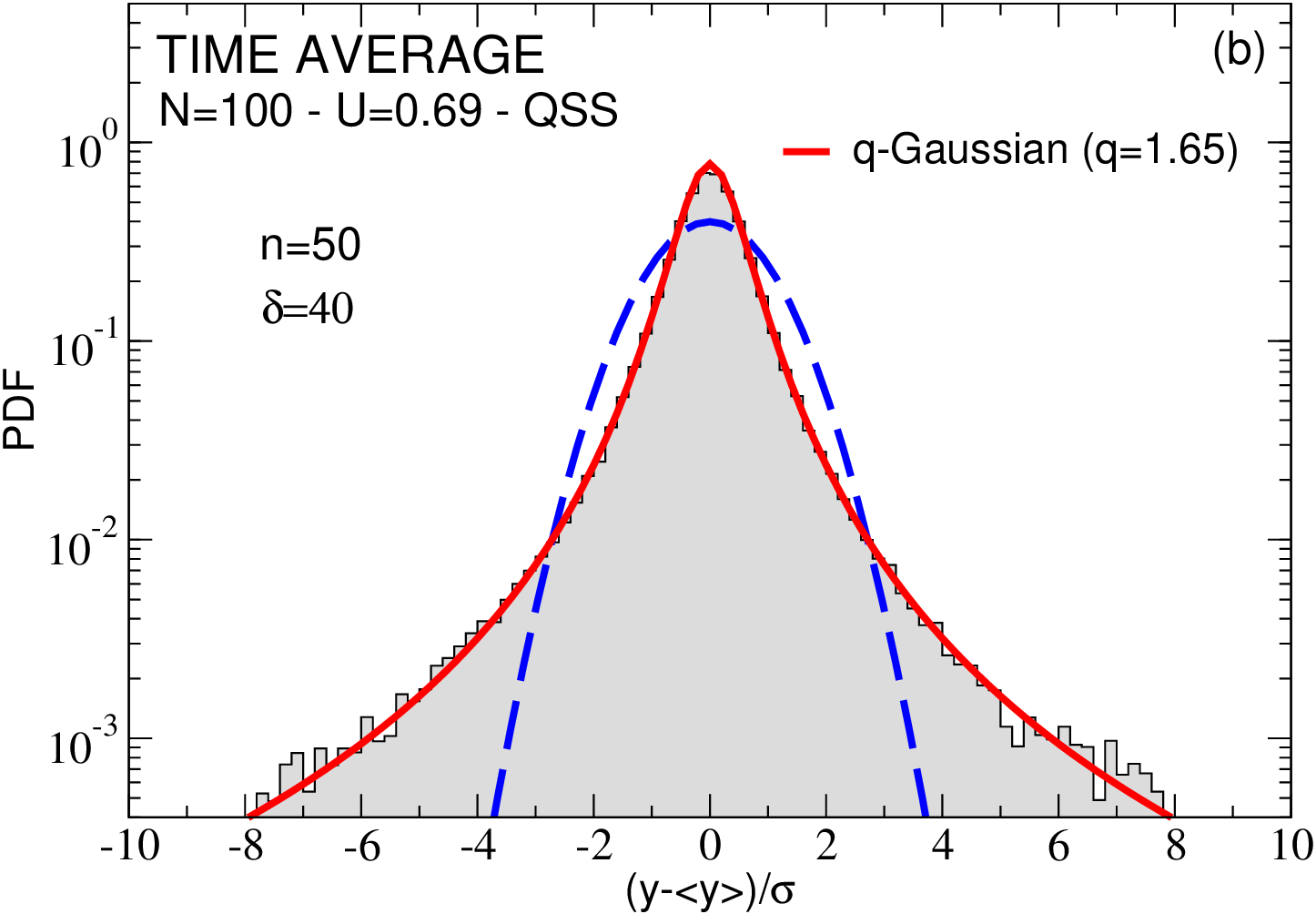}
\caption{Numerical simulations for the HMF model, with $N=100$ and $U=0.69$ and M1 initial conditions. We are in this case \textit{inside the QSS regime}. (a) We plot here the Pdf  of the single rotor velocities at  time $t=100$ (ensemble average over 1000 realizations). The shape is not Gaussian. (b) Time average Pdf  for the  normalized variable  $y$ with $n=50$ and with a time interval $\delta=40$, calculated after a  transient time of 100  time units. An average over 1000 different realizations of the initial conditions was also considered in order to have a good statistics. The resulting shape is very different from that one shown in (a) and can be very well fitted with a $q$-Gaussian. 
}
\label{fig3}
\end{center}
\end{figure}
%

\begin{figure}[h]
\begin{center}
\includegraphics[scale=0.17]{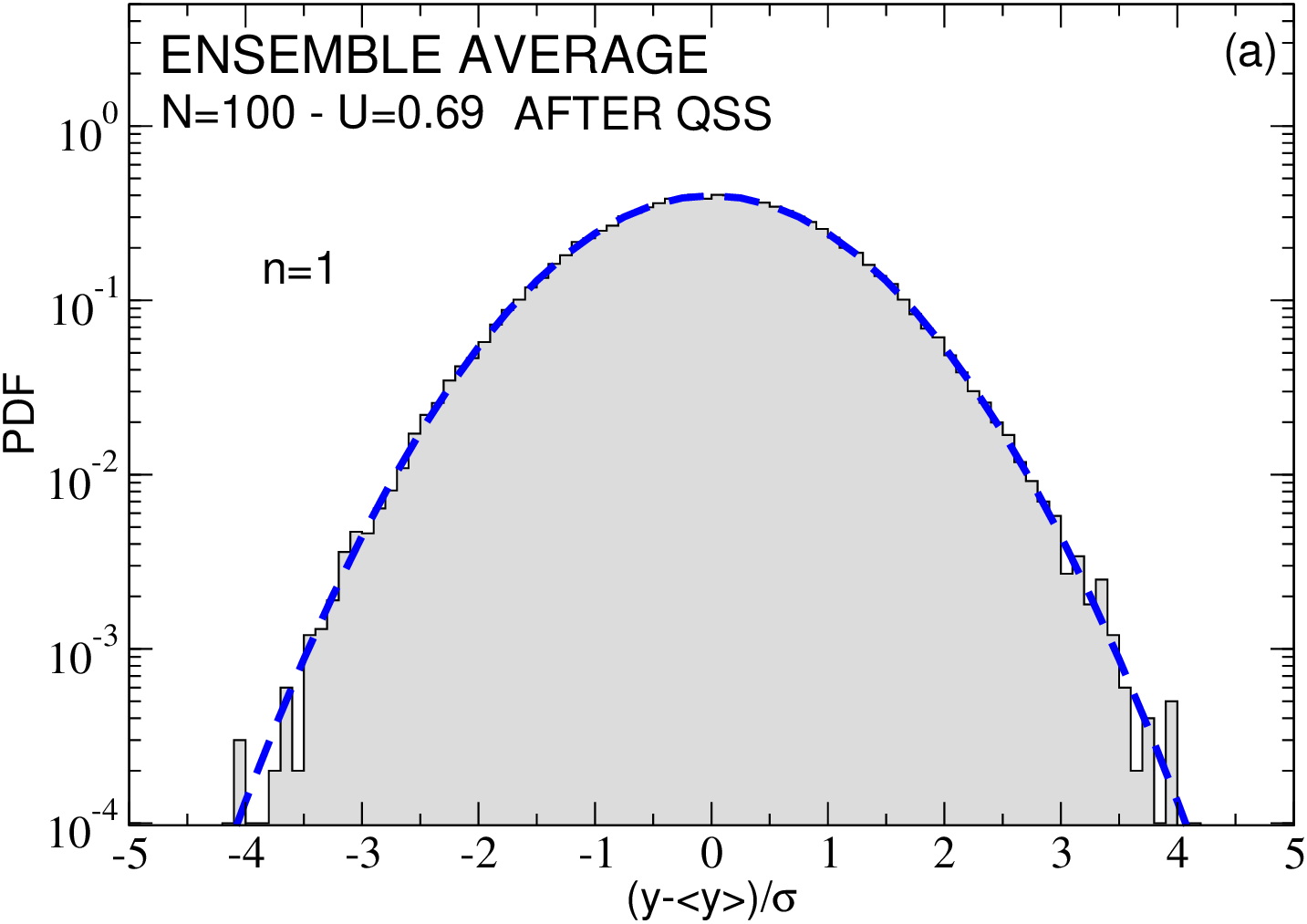}
\includegraphics[scale=0.17]{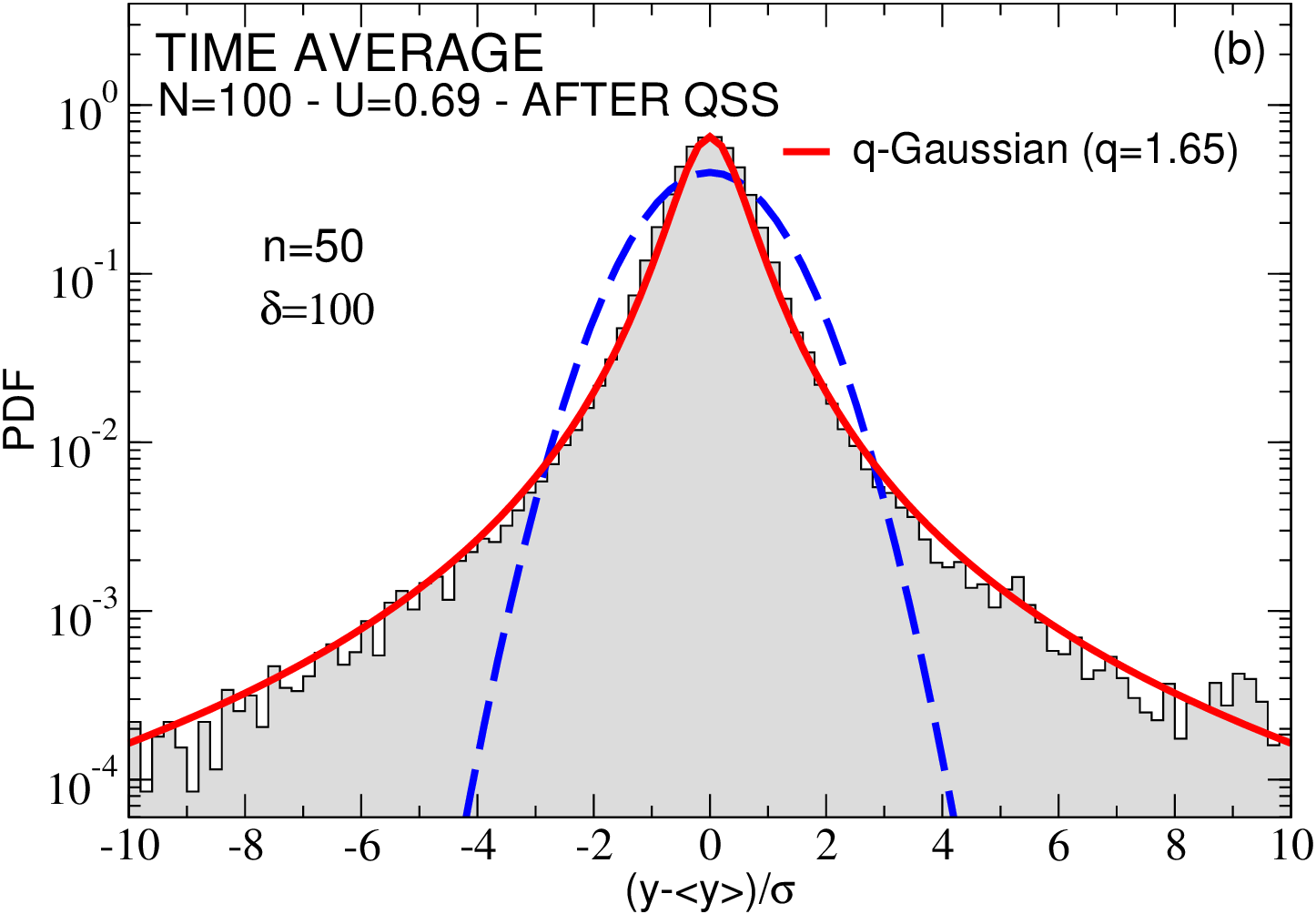}
\includegraphics[scale=0.17]{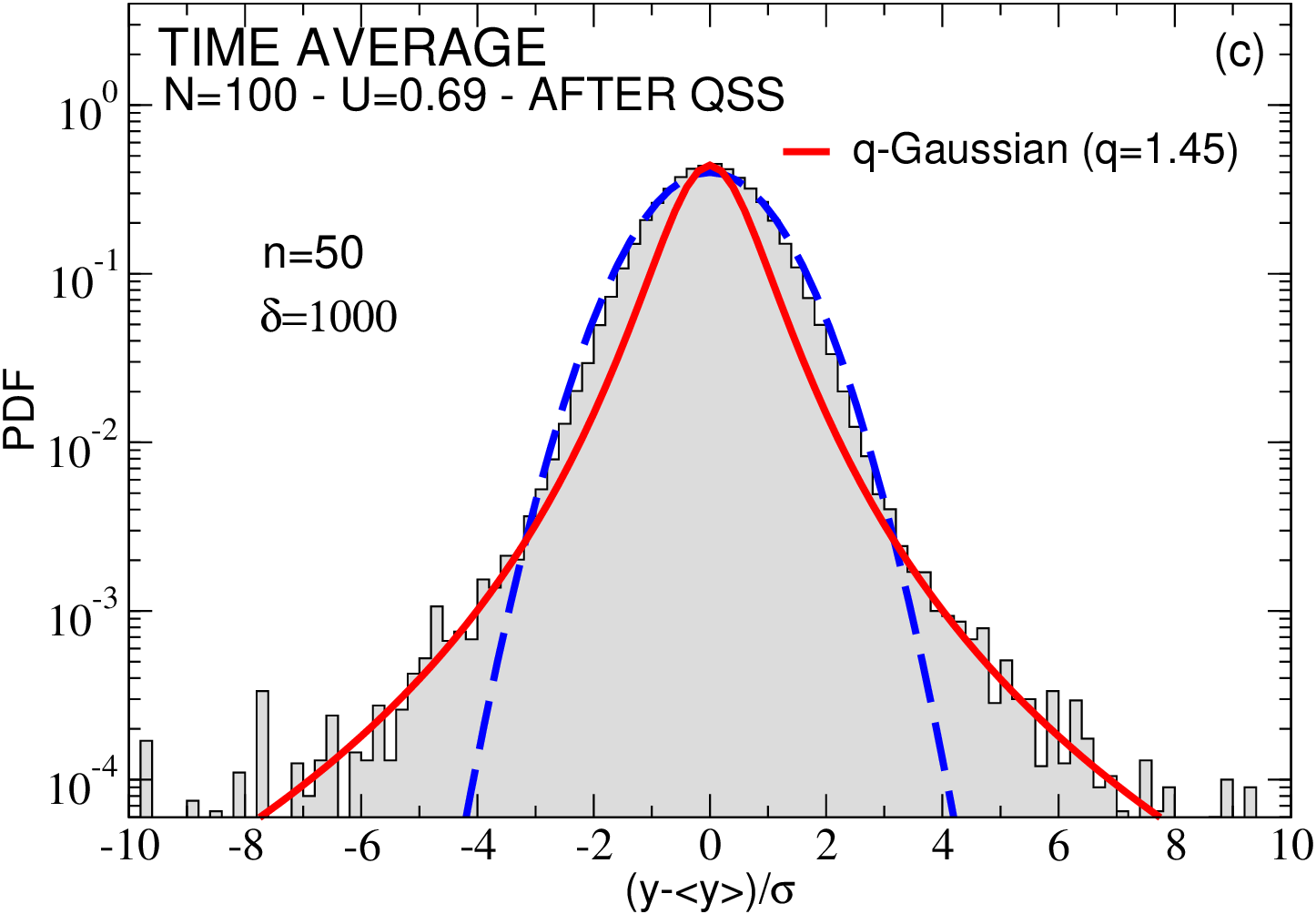}
\includegraphics[scale=0.17]{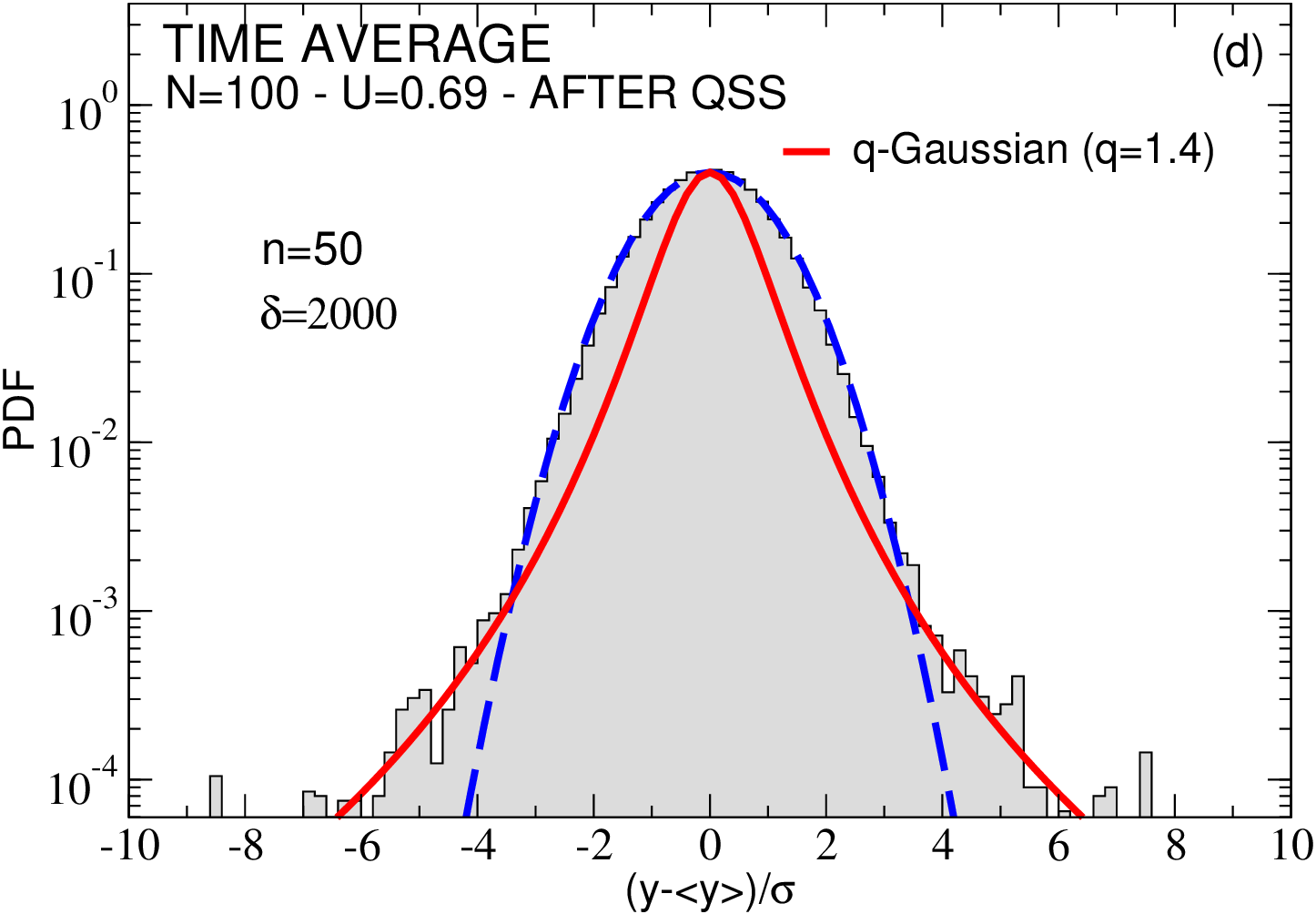}
\includegraphics[scale=0.17]{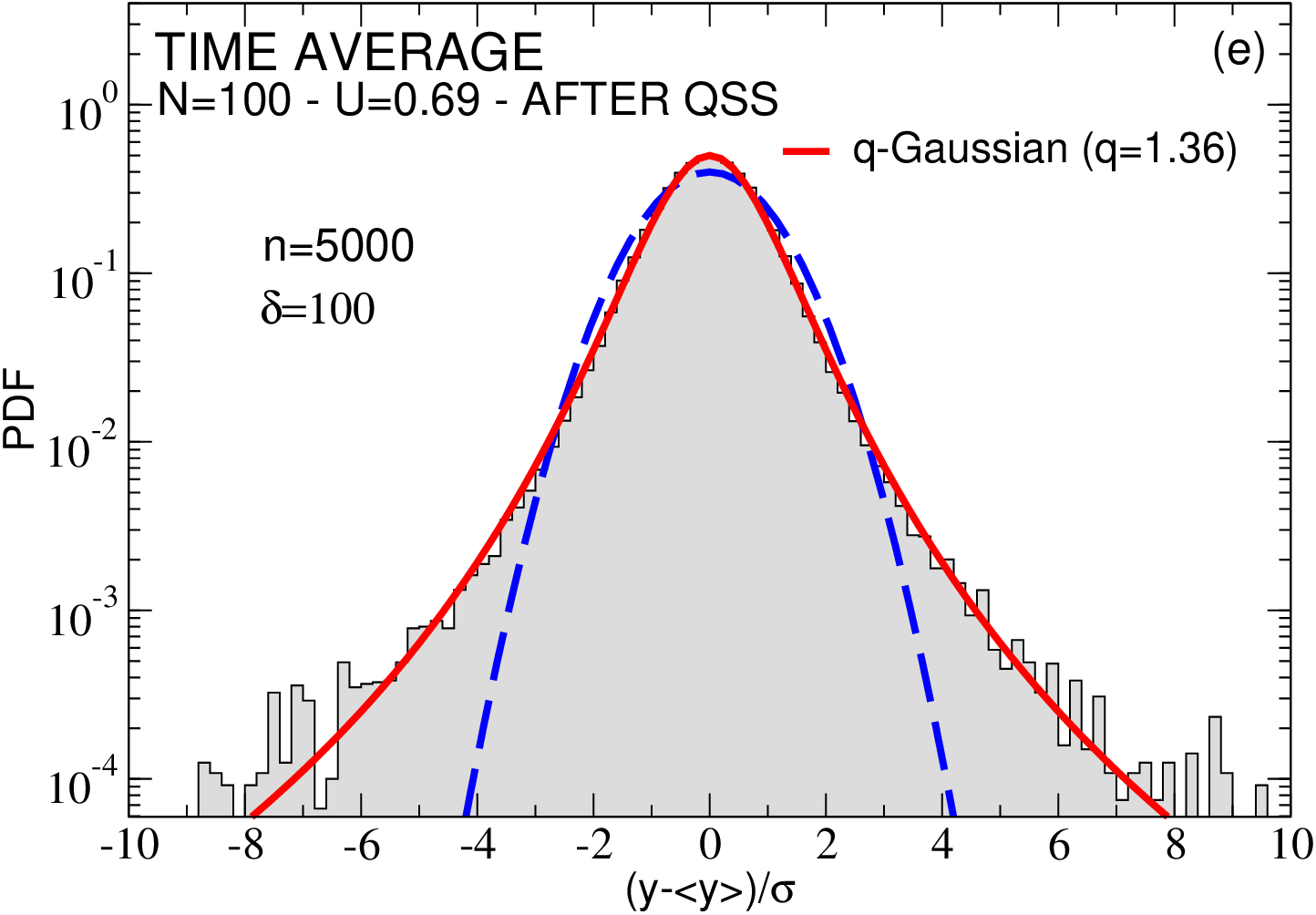}
\includegraphics[scale=0.17]{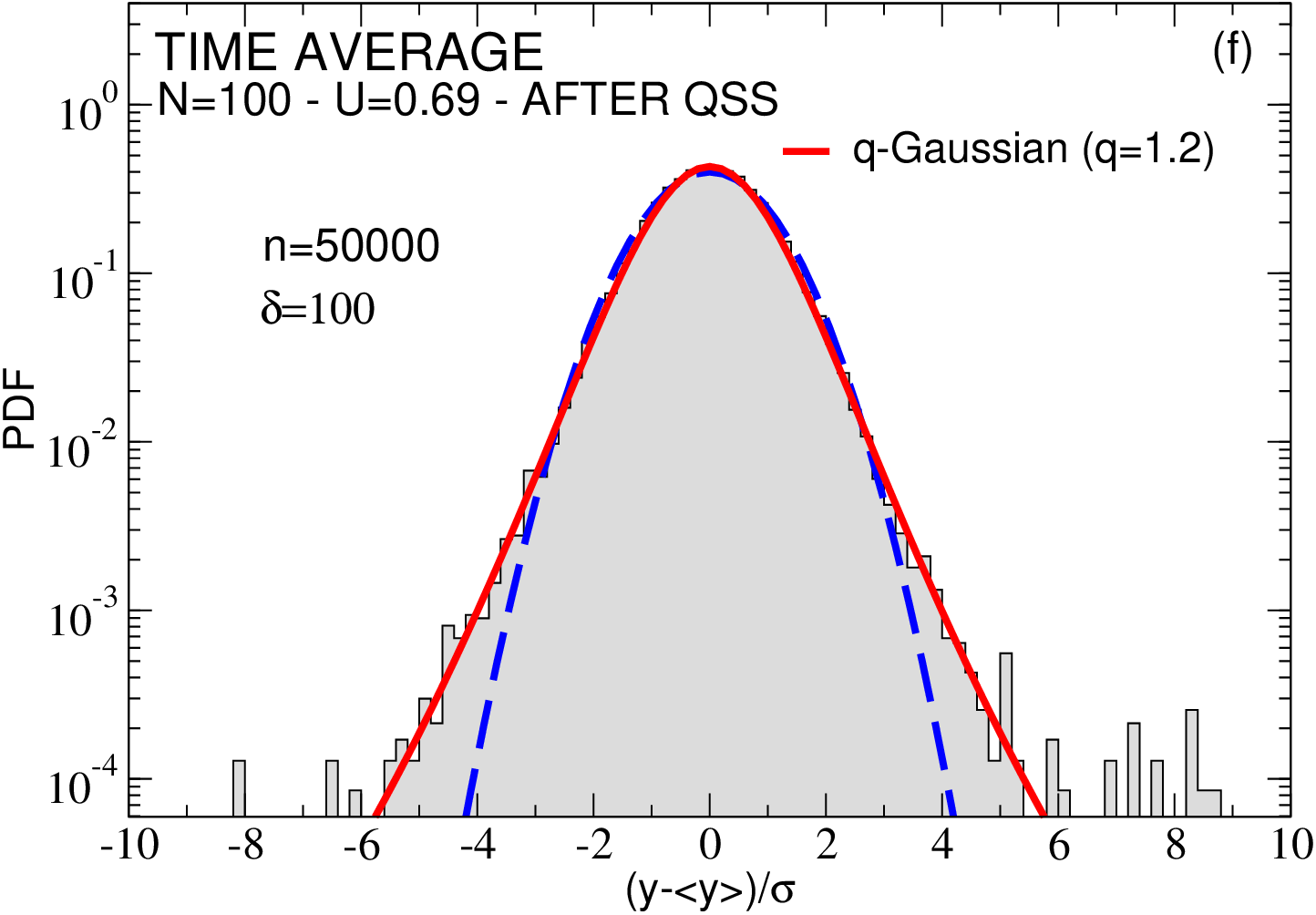}
\caption{Numerical simulations for the HMF model, with $N=100$, $U=0.69$  and M1 initial conditions. We are here \textit{after the QSS regime}. (a) 
Pdf of the single rotor velocities at t=40000 (ensemble average over 1000  realizations). The shape is Gaussian since the system is at equilibrium. In the other figures we plot the time average Pdfs for the  normalized variable  $y$, calculated after a transient time of $40000$. An average over 1000 different realizations of the initial conditions was also considered in order to have a good statistics. In figs.(b-d) we considered  $n=50$ but with different time intervals, more precisely $\delta$=100 (b), $\delta$=1000 (c) and $\delta$=2000 (d), over an increasing simulation time at equilibrium.  
(e)-(f) 
Pdfs obtained  by keeping fixed the value $\delta =100$ and increasing  the number $n$ of velocities  in the  sum for getting $y$.  More  precisely, $n=5000$ in (e) and $n=50000$ in (f). It is clear that, both for $\delta\rightarrow \infty$ and $n\rightarrow \infty$, the Pdfs shape tends to a Gaussian.
 }
\label{fig4}
\end{center}
\end{figure}
\subsection{N=100 and U=0.69}
Let us to consider now numerical simulations for the HMF model with size $N=100$  and  $U=0.69$. In this case a QSS regime exists, but its characteristic lifetime is quite short  since the  noise induced by the  finite size  drives the system  towards equilibration rapidly.  However  strong correlations, created  by the $M1$ initial conditions, exist and their decay is  slower than in the case $U=0.4$. In fig. \ref{fig3} we show in (a) the  Pdf of the velocities calculated at $t=100$ (i.e. at the beginning of the QSS regime). An ensemble average  over 1000 realizations was considered. The  Pdf shows a strange  shape which remains constant in the QSS, as already observed in the  past \cite{hmf2}, and which differs from both the Gaussian and the $q$-Gaussian curves.
On the other hand, we show in (b) the Pdf of the  variable $y$ with $n=50$ and $\delta=40$, i.e. calculated over a total of $2000$ time steps after a transient of $100$ units, in order to stay inside the QSS temperature plateaux (see fig.\ref{fig1}).
In this case the system is weakly chaotic and non ergodic \cite{pluchino,ep1} and  the  numerical Pdf is reproduced very well by a $q$-Gaussian with $q=1.65\pm0.05$. Although in this case we have used differents initial conditions also for time averages, these results provide a first indication  that ensemble and time averages are inequivalent in the QSS regime. Note that, due to the shortness of the QSS plateaux, for $N=100$ it is not possible to use greater values of $\delta$ or $n$ in the numerical calculations of the $y$'s.

In fig.\ref{fig4} we repeat the previous simulations for $N=100$ and $U=0.69$, but adopting a transient time of 40000 steps, in order to study the behavior of the system \textit{after} the QSS regime.
The ensemble average Pdf (over 1000 realizations) of the single rotor velocities at the time $t=40000$ is shown in (a) and indicates that equilibrium seems to have been reached. In fact the  agreement with the standard  Gaussian is almost perfect up to $10^{-4}$. In the other figures we plot the time average Pdfs for the variable $y$ with $n=50$ and for different time intervals $\delta$, as done for $U=0.4$.  More precisely $\delta$=100 in (b), $\delta$=1000 in (c) and $\delta$=2000 in (d). Again it is evident a strong dependence of the Pdf shapes  on  the time interval $\delta$ adopted. In fact initially (b) the Pdf is well fitted by a $q$-Gaussian with a  $q=1.65\pm0.05$, however increasing  $\delta$, in (c) and (d), the central part of the Pdf becomes Gaussian while tails are still present and can be well fitted by $q$-Gaussians with values of $q$ that tend towards unity. However, at variance with the $U=0.4$ case, in this case not even a time interval $\delta=2000$ is sufficient to reach a complete Gaussian-shaped Pdf down to $10^{-4}$: evidently the strong correlations characterizing the QSS regime decay very slowly even after it, making the equilibrium shown by the ensemble average Pdf in (a) only apparent. This means that full ergodicity, i.e.,  full equivalence between ensemble and time averages, is reached, in this case, only asymptotically.

The last statements are confirmed by panels (e) and (f) of fig.\ref{fig4}, where the effect of increasing the  number $n$
of summed velocities, keeping fixed  the  value  of $\delta$, has been investigated.
More  precisely $\delta$=100 and $n=5000$ in (e) and $n=50000$ in (f).  As expected, the increment of $n$ makes the Pdf closer to the Gaussian, essentially because  the  total time  over which the sum is considered increases (for $n=50000$ we cover a simulation time of $5\times 10^6$) and therefore correlations become asymptotically weaker and weaker, thus finally
satisfying the prediction of the standard CLT

In order to study in more details the ensemble-time inequivalence along the QSS regime in the next subsection we will increase the system size and discuss numerical results for $N=5000$ and $N=50000$.

\begin{figure}
\begin{center}
\includegraphics[scale=0.23] {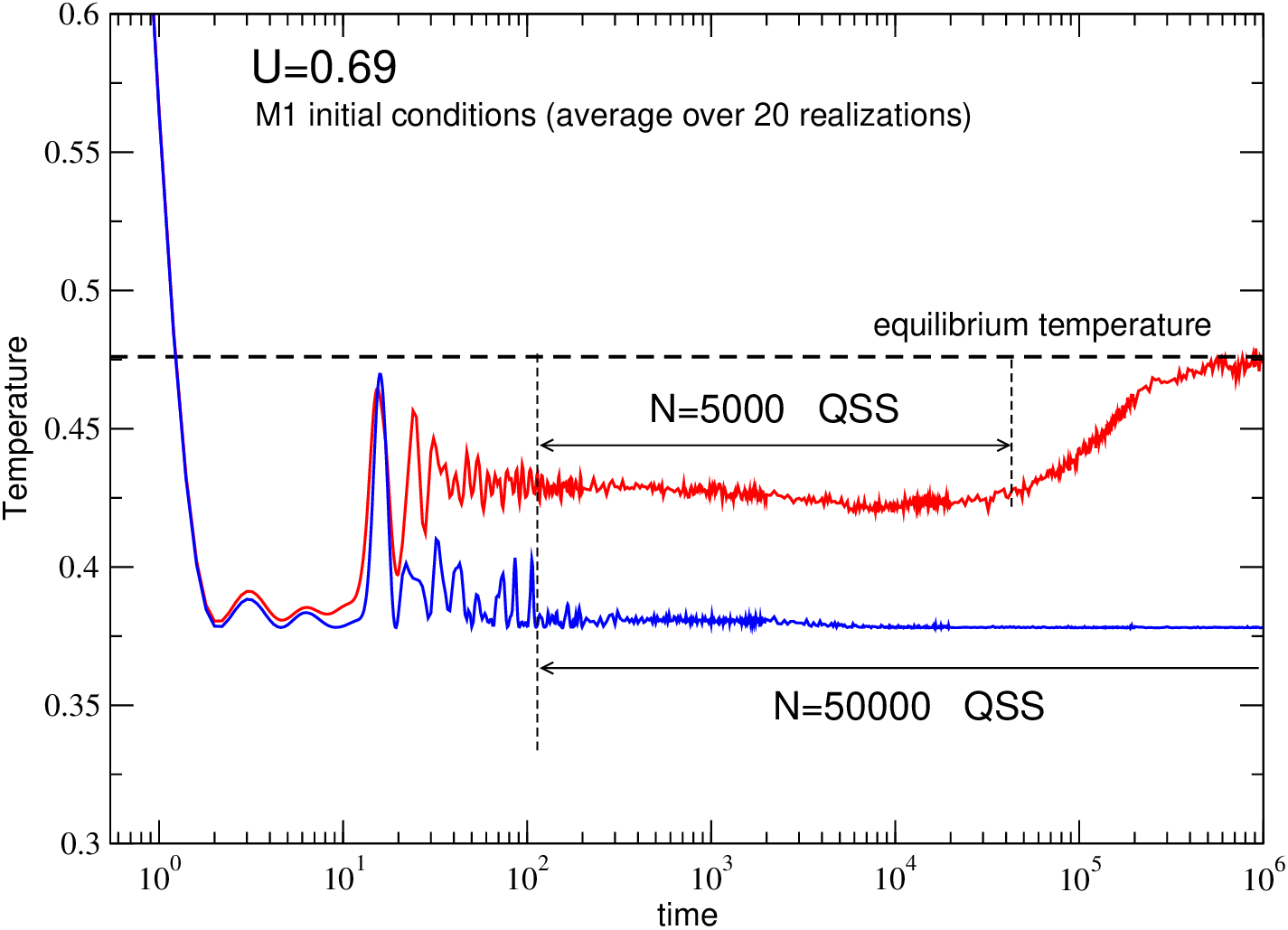}
\caption{Temperature time evolution for the HMF system, with $U=0.69$, M1 initial conditions and for $N=5000$  and $N=50000$.
The presence of a long-lasting QSS regime is clearly visible in both the cases and the plateaux are very much larger then in the $N=100$ case.}
\label{fig5}
\end{center}
\end{figure}

\begin{figure}[h]
\begin{center}
\includegraphics[scale=0.17]{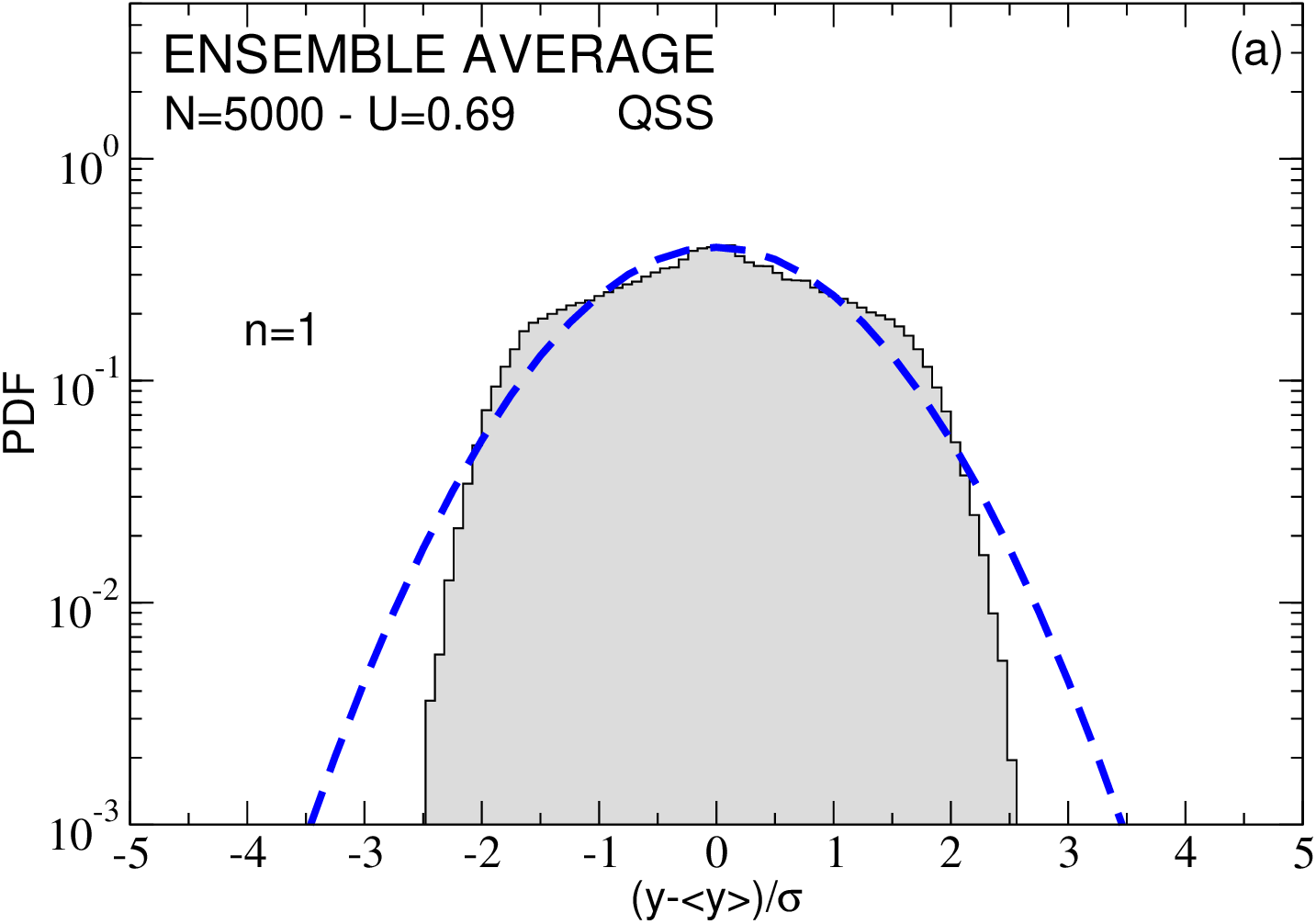}
\includegraphics[scale=0.17]{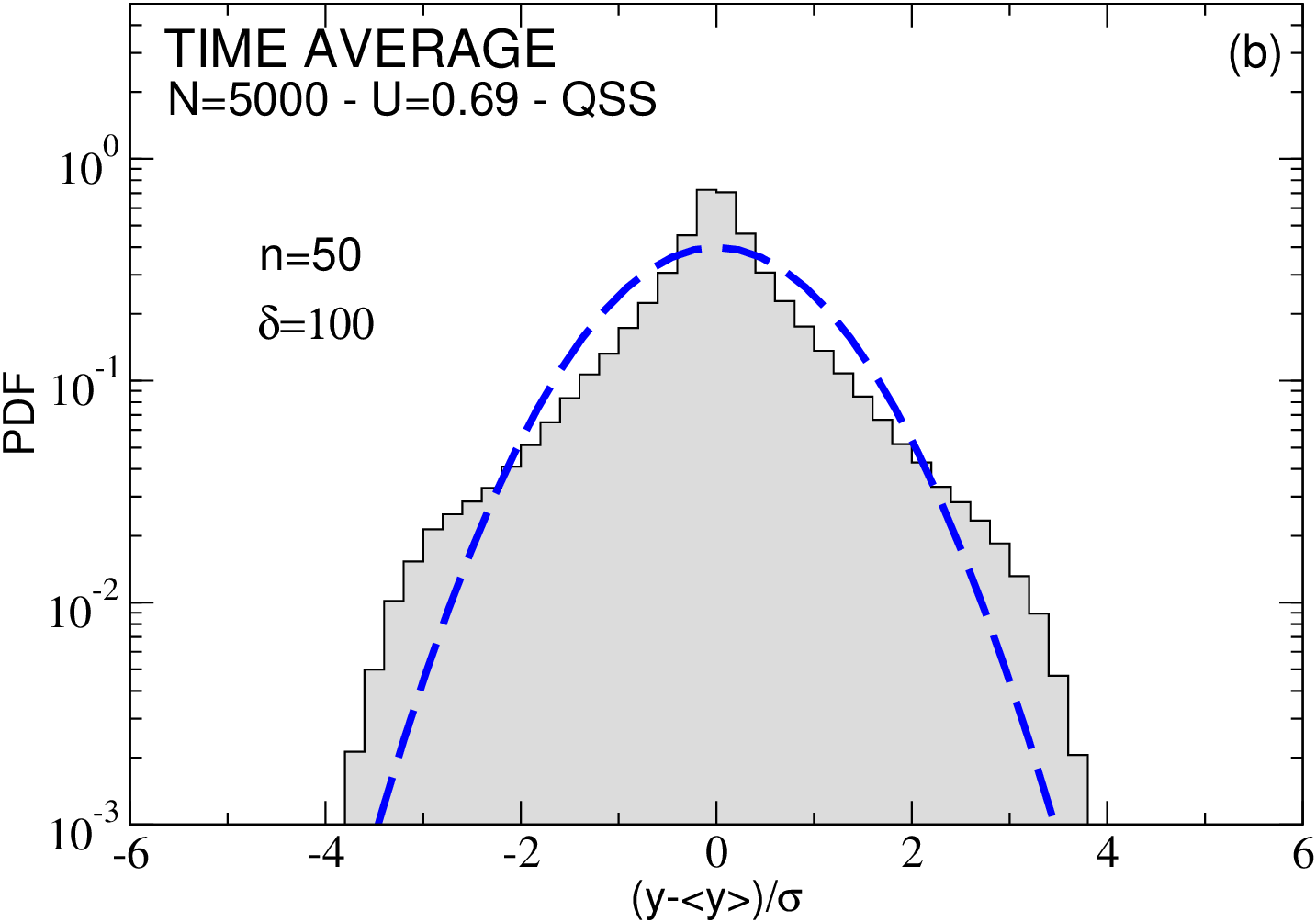}
\includegraphics[scale=0.17]{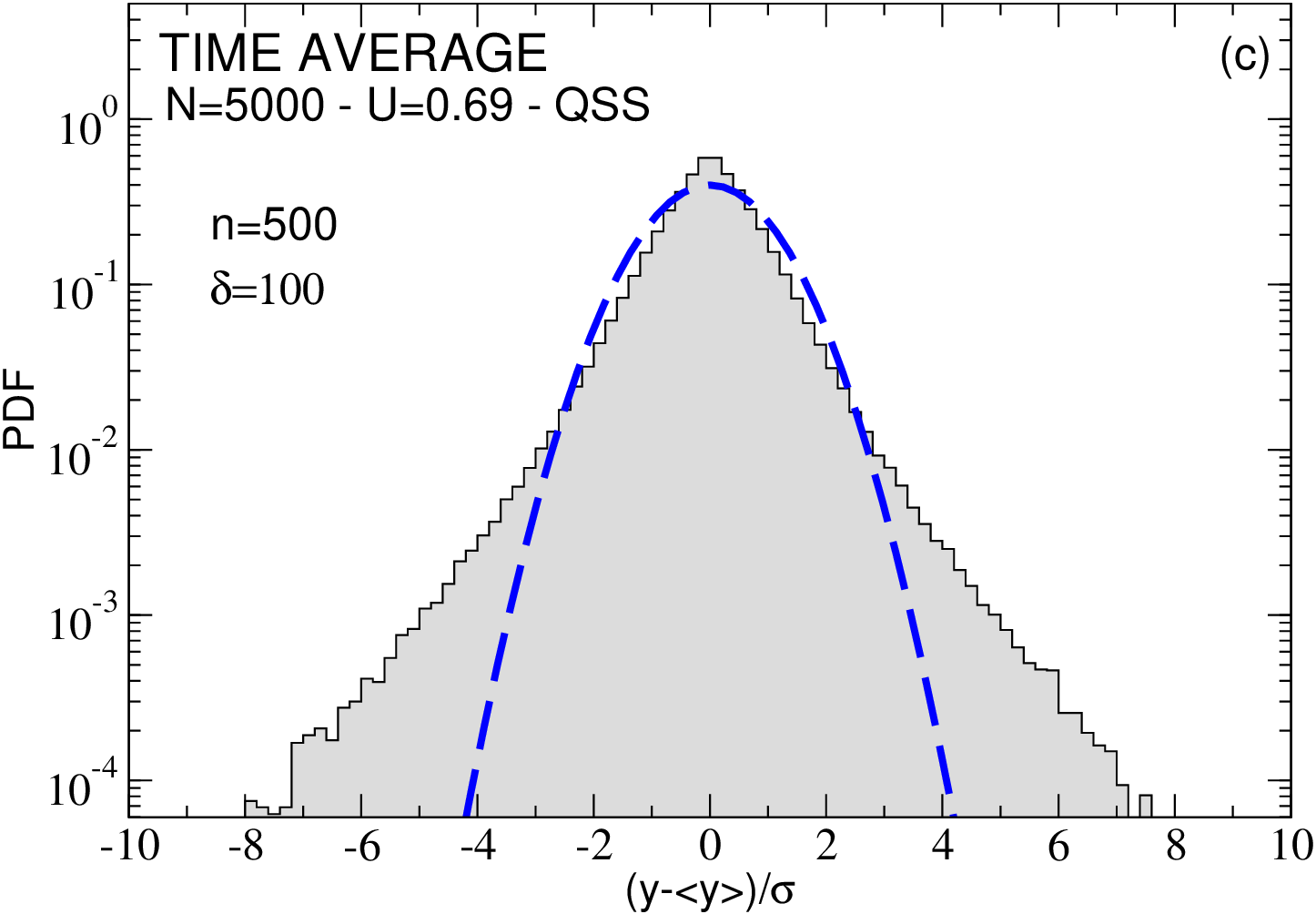}
\includegraphics[scale=0.17]{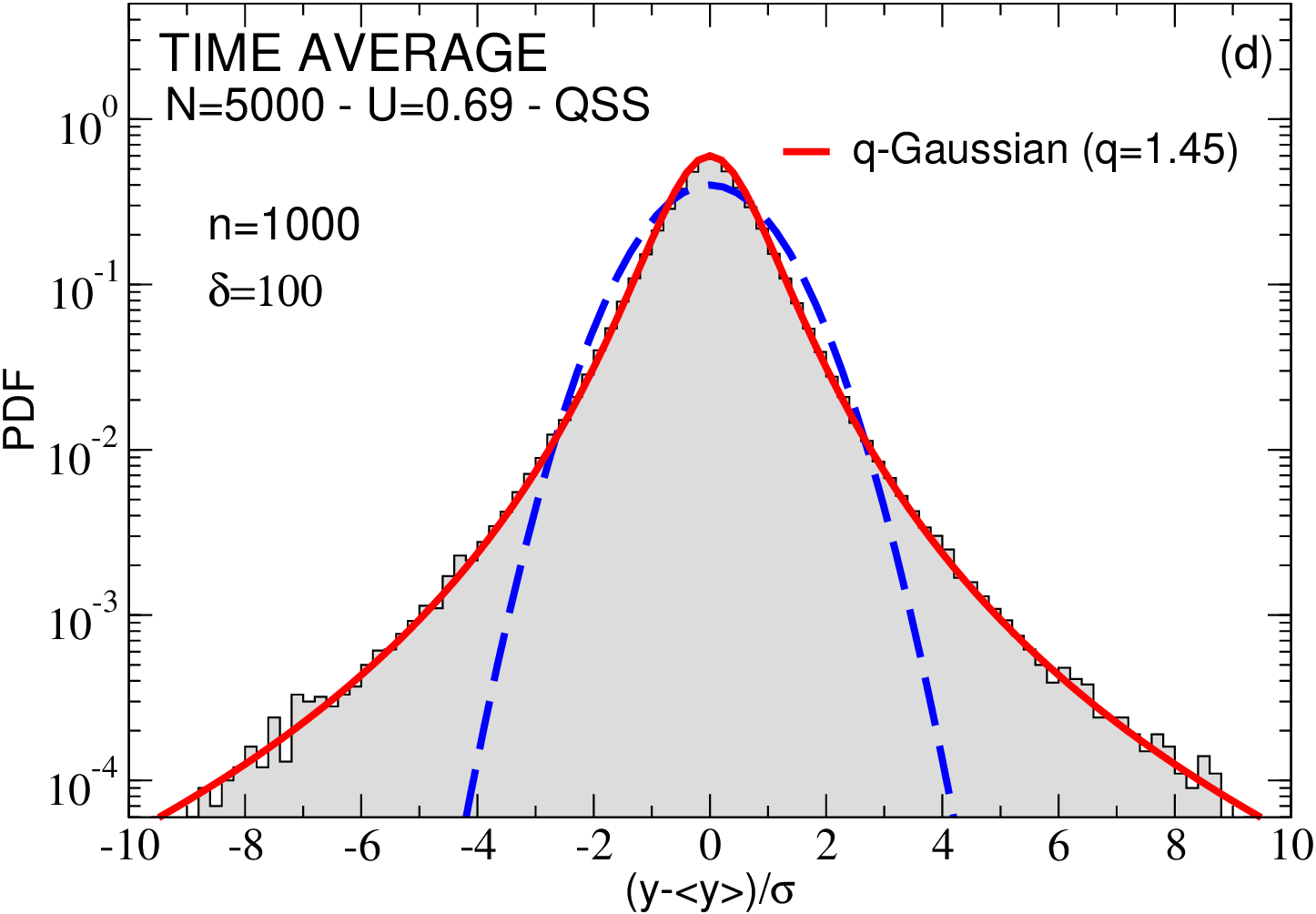}
\caption{ Numerical simulations for the HMF model with $N=5000$, $U=0.69$ and M1 initial conditions, \textit{in the QSS regime}. (a) 
Pdf  of single rotor velocities at the time $t=100$ (ensemble average over 1000 realizations). (b-d) 
Time average Pdfs for the normalized variable $y$, after a transient time of $100$ units and considering increasing values of $n$ with a fixed time interval $\delta=100$, i.e. considering an increasing total simulation time inside the QSS. An average over $200$ different realizations of the initial conditions was also considered in order to have a good statistics. Only for $n=1000$, i.e. when the entire QSS extension has been considered (see fig.\ref{fig5}), we get a very good $q$-Gaussian shape. 
}
\label{fig6}
\end{center}
\end{figure}

\subsection{N=5000 and N=50000 at U=0.69}

In fig.\ref{fig5} we show  the  time evolution of the temperature for the cases $N=5000$ and $N=50000$ at $U=0.69$, always starting (as usual) from the M1 initial conditions.
It is evident that, for both systems, the length of the QSS plateaux is very much greater than for $N=100$.

We discuss first numerical simulations done inside the QSS for $N=5000$ and $U=0.69$.

In fig.\ref{fig6} we show in (a) the ensemble average Pdf of velocities calculated over $1000$ realizations
at $t=100$, i.e. at the beginning of the QSS regime.
Its shape, constant along the entire  QSS,  is clearly not Gaussian and looks similar to that of fig.\ref{fig3} (a).
In panels (b-d) we show the effect of increasing the number $n$ of velocity terms in the $y$ sum on the time average Pdfs, calculated using a fixed value of $\delta=100$. An average over 200 different realizations of the initial conditions was also considered in order to have good statistics. In this case only for $n=1000$ a $q$-Gaussian, with $q=1.45\pm0.05$, emerges. This is most likely due not to the effective number of $n$ used but, consistently with
fig.\ref{fig6}, to the  fact that  when choosing a large $n$ one is averaging over a larger interval of time
and thus  considers in a more appropriate  way the  average  over the entire QSS  regime.
In any case the observed behavior goes in the opposite direction to the prescriptions of the standard CLT and to the trend shown in panels (e-f) of fig.\ref{fig4}. Indeed, increasing $n$, the Pdf tails do not vanish but become more and more evident, thus  supporting even further the claim about the existence of a non-Gaussian attractor for the nonergodic QSS regime of the HMF model.
Moreover, the results of fig.\ref{fig6} confirm the robustness of the $q$-Gaussian shape along the entire QSS plateaux and the inequivalence between ensemble and time averages in the metastable regime.

\begin{figure}
\begin{center}
\includegraphics[scale=0.178]{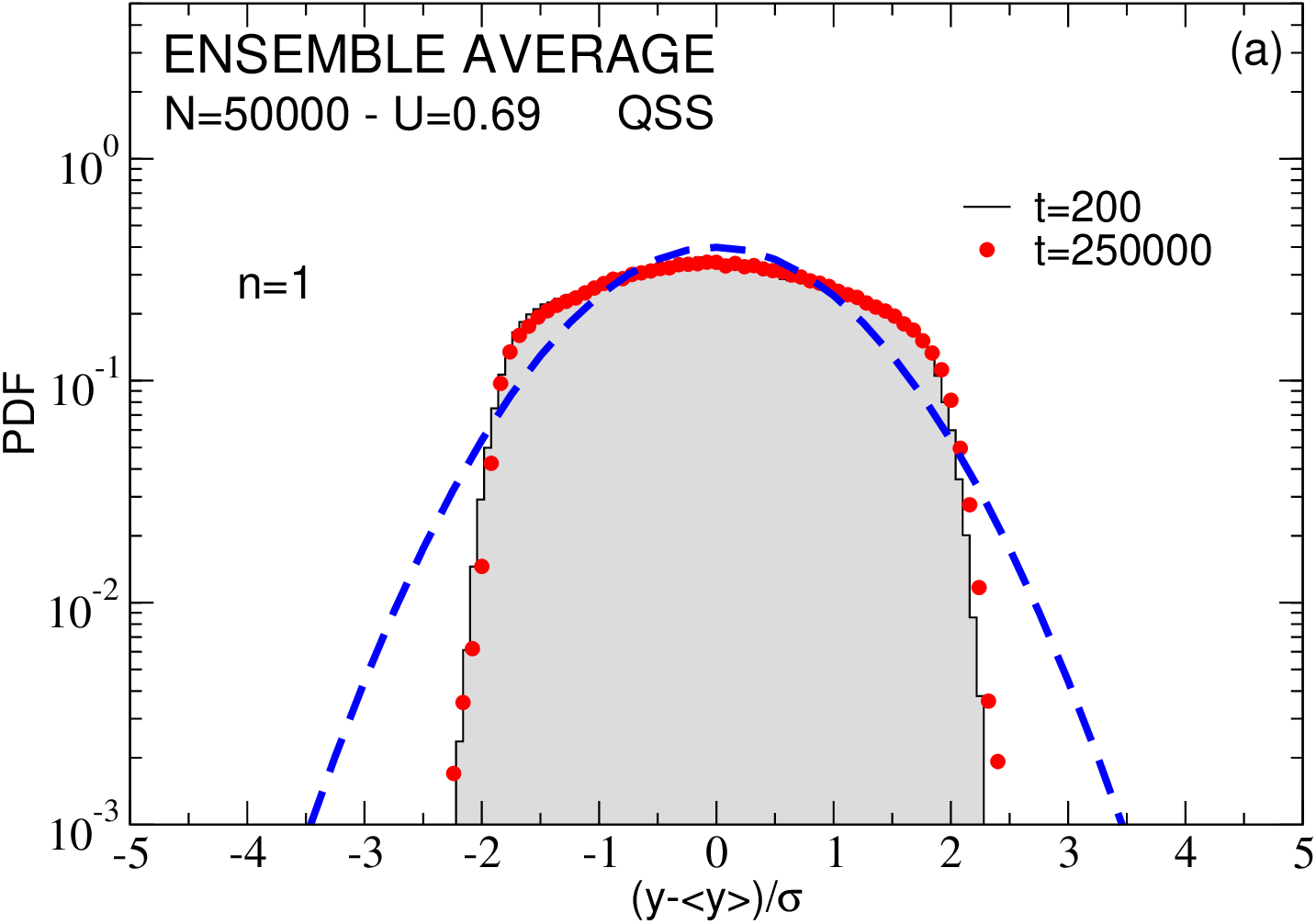}
\includegraphics[scale=0.178]{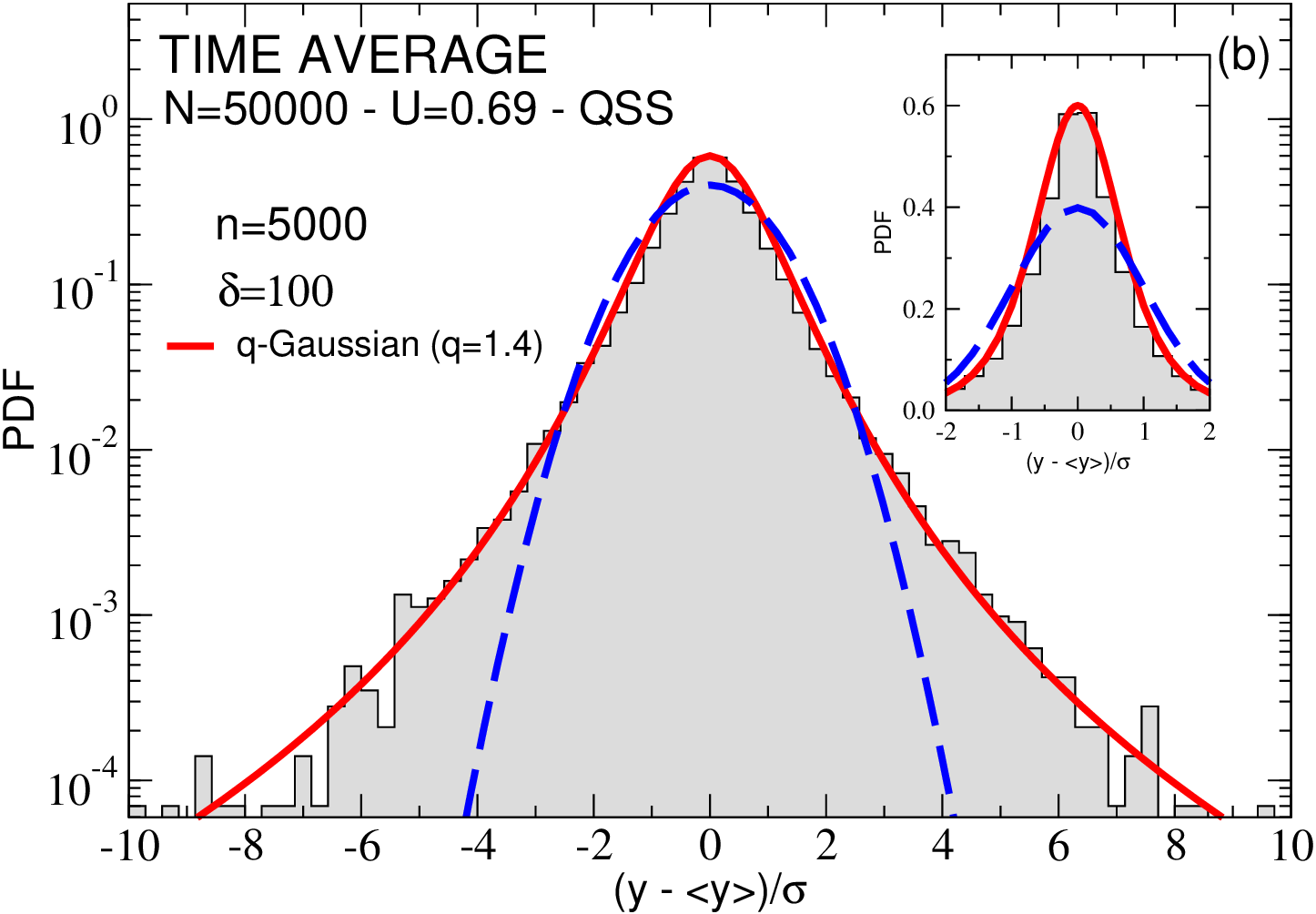}
\caption{ Numerical simulations for the HMF model for N=50000, U=0.69 and M1 initial conditions \textit{in the QSS regime}. (a) 
Pdfs of single rotor velocities at the times t=200 and t=250000 (ensemble average over 100 realizations). (b) 
Time average Pdf for the  variable $y$ calculated over \textit{only one single realization} in the QSS regime and after a transient time of 200 units. In this case we used $\delta=100$ and $n=5000$, in order to cover a very large portion of the QSS (see fig.\ref{fig5}). Again, a  $q$-Gaussian reproduces very well the  calculated Pdf both in the tails and in the central part (see inset). See text for further details.}
\label{fig7}
\end{center}
\end{figure}

Let us now definitively demonstrate this inequivalence considering the case N=50000 at U=0.69.
In fig.\ref{fig7} (a) we plot the ensemble average Pdf of the velocities calculated (over 100 different realizations) at $t=200$, i.e. at the beginning of the QSS regime, and after a very long transient, at $t=250000$ (full circles). In panel (b) we plot the time average Pdf for the normalized variable $y$ with $n=5000$ and $\delta=100$, after a transient of $200$ time units and over a simulation time of $500000$ units along the QSS.
It is important to stress that in this case \textit{only one single realization} of the initial conditions has been performed, realizing this way a \textsl{pure time average}.
The shape of the time average Pdf (b) results to be again  a robust $q$-Gaussian, with $q=1.4\pm0.05$,
not only in the tails, but also in the center (see inset).
The time average Pdf is completely different from  the ensemble average Pdf of fig.\ref{fig7}(a) (that is also very robust over all the plateaux), thus  confirming definitively the inequivalence between the two kind of averages and the existence of a $q$-Gaussian attractor in the QSS regime of the HMF model.  These results indicate  that  standard
statistical mechanics  based  on the ergodic hypothesis  cannot be applied in this case, while  a generalized version, like the $q$-statistics \cite{tsallis0,tsallis} is likely more suitable \cite{ep1}.   
%

\section{Conclusions}

The numerical simulations presented in this paper strongly indicate that dynamical correlations and ergodicity breaking, induced in the HMF model by the initial out-of equilibrium violent relaxation,  are present along the entire QSS metastable regime and decay very slowly even after it.  In particular,  considering finite  sums  of $n$ correlated  variables  (velocities in this case) selected with a constant time interval $\delta$ along single rotor trajectories, allowed us  to  study this phenomenon in detail. Indeed, we numerically showed that, in the weakly chaotic QSS regime, (i) ensemble average and time average of velocities are inequivalent, hence the ergodic hypothesis is violated, (ii) the standard CLT is violated, and (iii) robust $q$-Gaussian attractors emerge.
On the contrary,  when  no  QSS exist, or at a very large time after equilibration, i.e., when the system is fully chaotic and ergodicity has been restored,  the ensemble average of velocities results to be equivalent to the time average and  one  observes a convergence  towards the standard Gaussian attractor. In this case,  the  predictions  of CLT are satisfied, even if we have only considered a finite sum  of stochastic variables. How fast this happens   depends on the size $N$, on the number $n$ of terms summed in the $y$ variables and on the time interval $\delta$ considered.

These results are consistent with the recent $q$-generalized forms of the CLT discussed in the literature  \cite{tsallisMilan,umarov,tirnakli,comment,plastino}, and pose severe questions  to the often adopted  procedure  of using ensemble averages   instead of time averages.  Nonergodicity in coupled many particle systems goes back to the famous FPU experiment \cite{fpu}, but in our case is due to the long-range nature of the interaction.
 More recently, nonergodicity was found in deterministic iterative systems  exibiting subdiffusion \cite{barkai}, but also in real experiments of  shear flows, with results that were fitted with Lorentzians, i.e., $q$-Gaussians with $q=2$ \cite{wang}. 
The whole scenario reminds  that found for  the  leptokurtic  \textsl{returns} Pdf   in financial markets \cite{mantegna}, or in turbulence \cite{beck}, among many other systems, 
and   could   probably  explain  why  $q$-Gaussians appear to be ubiquitous  in complex systems.  
Finally, we would like to add that, although it is certainly nontrivial to prove analytically whether the  attractor in the nonergodic QSS regime of the HMF model  precisely is a $q$-Gaussian or not (analytical results, as well as numerical dangers, have been recently illustrated in ref.\cite{hilhorst} for various models), our numerical simulations unambiguously provide a very strong indication towards  the existence of a robust $q$-Gaussian attractor in the case considered. This  opens new ways to the possible application of the $q$-generalized statistics in long-range Hamiltonian systems which will be explored in future papers.

\acknowledgments
We  thank  Marcello Iacono Manno for many technical discussions and help in the preparation 
of the  scripts  to run  our   codes on the GRID platform. The  numerical calculations here presented 
were done within the TRIGRID project.
A.P. and A.R. acknowledge financial support from the PRIN05-MIUR project "Dynamics and Thermodynamics 
of Systems with Long-Range Interactions". C.T. acknowledges financial support from the Brazilian 
Agencies Pronex/MCT, CNPq and Faperj.

%

\end{document}